\documentclass[graybox]{svmult}

\usepackage{mathptmx}       
\usepackage{helvet}         
\usepackage{courier}        
\usepackage{graphicx}        
\usepackage{multicol}        

\begin{document}

\title*{A finite element approach of the behaviour of woven materials at microscopic scale}
\author{Damien Durville}
\institute{Damien Durville \at LMSSMat - Ecole Centrale Paris , Grande Voie des Vignes, 92290 Chatenay-Malabry, France, \email{damien.durville@ecp.fr}}
%
%
\maketitle
\abstract*{A finite element simulation of the mechanical behaviour of woven textile materials at the
  scale of individual fibers is proposed in this paper. The aim of the simulation is to understand
  and identify phenomena involved at different scales in such materials. The approach considers
  small patches of woven textile materials as collections of fibers. Fibers are modelled by 3D beam
  elements, and contact-friction interactions are considered between them. An original method for
  the detection of contacts, and the use of efficient algorithms to solve the nonlinearities of the
  problem, allow to handle patches made of few hundreds of fibers. The computation of the unknown
  initial configuration of the woven structure is carried out by simulating the weaving process.
  Various loading cases can then be applied to the studied patches to identify their mechanical
  characteristics.  To approach the mesoscopic behaviour of yarns, 3D strains are calculated at the
  scale of yarns, as post-processing. These strains display strong inhomogeinities, which raises the
  question of using continuous models at the scale of yarns.}

\abstract{A finite element simulation of the mechanical behaviour of woven textile materials at the
  scale of individual fibers is proposed in this paper. The aim of the simulation is to understand
  and identify phenomena involved at different scales in such materials. The approach considers
  small patches of woven textile materials as collections of fibers. Fibers are modelled by 3D beam
  elements, and contact-friction interactions are considered between them. An original method for
  the detection of contacts, and the use of efficient algorithms to solve the nonlinearities of the
  problem, allow to handle patches made of few hundreds of fibers. The computation of the unknown
  initial configuration of the woven structure is carried out by simulating the weaving process.
  Various loading cases can then be applied to the studied patches to identify their mechanical
  characteristics.  To approach the mesoscopic behaviour of yarns, 3D strains are calculated at the
  scale of yarns, as post-processing. These strains display strong inhomogeinities, which raises the
  question of using continuous models at the scale of yarns.}

\section{Introduction}

The mechanics of woven materials can be adressed at three different scales : the macroscopic scale
relevant to pieces of fabric, the mesoscopic scale related to yarns, and the microscopic scale
concerning fibers inside yarns. Different approaches are available to handle the problem of the
identification and characterization of the complex mechanical behaviour of these materials. Some
developments are specially dedicated to the characterization of the geometry of yarns in the woven
structure (\cite{Verpoest05,Lomov07}). Geometries of yarns obtained by these methods can then be
meshed and classical finite element codes can be employed to compute the mechanical response of
textile or textile composite structures. Other approaches concentrate on the finite element
simulation at the scale of yarns, representing yarns by means of 3D elements with appropriate
constitutive laws \cite{Bois05}. Going down to the scale of fibers requires a description of the
geometry of fibers in the initial configuration. Such a geometry is a priori unknown, and therefore
needs to be computed, for example by simulating the weaving process. Such an approach has been
proposed using a finite element simulation code based on an explicit solver \cite{Finckh04}. The
method we present here is based on a simulation code specially developped to handle the mechanical
behaviour of entagled media \cite{Dur05}, and makes use of an implicit solver. This simulation code
is able to handle small patches of fabric made of few hundreds of fibers in order to identify their
mechanical behaviour \cite{Dur08}.  After recalling some basic features of the method, the way the
weaving process is simulated to compute the initial configuration of the woven structure is
presented, and results for two different weave patterns are given. Classical loading tests are then
applied to these patches. The last part is dedicated to the calculation of 3D strains at the scale
of yarns by means of meshes generated as post-processing. The results show strong inhomegeneities of
strains at the scale of yarns, induced by localized displacements taking place between fibers.

\section{Purposes of the simulation at microcopic scale}

The needs related to the modelling and simulation of textile materials are diversified. Formability
studies require an identification of the global macroscopic behaviour of fabrics, whereas the focus
will be put on the local behaviour of individual fibers if damage and fiber breaking phenomena are
to be investigated.

The coupling between phenomena at different scales makes the study of such materials complex. The two
main mechanisms ruling the behaviour of textile materials, namely the interweaving of yarns
constituting the fabric, and the entanglement of fibers within the yarns, occur at different scales
but must be considered simultaneaously. One possible strategy to solve the problem at the
macroscopic scale is to formulate a mesoscopic model for yarns accounting for the behaviour of the
bundles of fibers making up the yarns. However, at the scale of yarns, one important issue is to
determine whether phenomena between discrete fibers within yarns may be approached by continuous
models.

To explore these questions, we propose to adress the global problem at the microscopic scale of
individual fibers, by considering small patches of fabrics as assemblies of individual fibers,
arranged in bundles, and developping between them contact-friction interactions. By this way, models
at intermediate scales are no longer required, and the only mechanical characteristics to be known
are those of individual fibers. As a downside, the initial geometry of fibers in the woven structure
can not be predicted a priori. It is therefore necessary to compute this initial geometry by
simulating the weaving process.

\section{Main features of the simulation code}

\subsection{A kinematically enriched beam model for fibers}

The beam elements used to take into account fibers in the simulation are based on an enriched
kinematical model that describes kinematics of each cross-section by the means of three vector
fields : one for the translation of the centroid, and two to represent planar and linear
deformations of the cross-section. This model, characterized by nine degrees of freedom, allows to
calculate full 3D strain tensors, accounting for deformations of cross-sections.

\subsection{Modelling of contact between fibers}

An original method \cite{Dur05,Dur08} has been developped to detect quicky and accurately the
numerous contacts taking place within a collection of fibers submitted to large deformations. This
method is based on the determination of contact elements made of pairs of material particles that
are predicted to enter into contact. The determination of these elements relies on the construction
of intermediate geometries in any region where two parts of beam are sufficiently close to each
other. These geometries are defined as the average of the two close parts of line. Normal directions
to these geometries are used as contact search directions to define material particles of contact
elements. By this way, the process of determination of contact elements is symmetrical with respect
to the two considered beams --which is not the case when the normal direction to only one structure
is chosen as contact search direction.

\subsection{Rigid bodies for the driving of boundary conditions}

To simulate the various conditions corresponding to the initial weaving process or the different
loading tests, a versatile driving of boundary considitions is required. For this purpose, rigid
bodies are attached to each end of yarn, and to each side of the patch. These rigid bodies can be
driven either by displacements or by forces, applied to their degrees of freedom in both translation
and rotation.

\section{Computation of the initial configuration : simulation of the weaving process}

\subsection{Modelling of the weaving process}

The simulation of the weaving process is necessary to determine the unknown initial configuration of the woven
structure. The weave pattern specifies which yarn must above or below the other at each crossing. The basic idea to
simulate the weaving process is to make these conditions progressively fulfiled.

To do so, we start from a flat configuration where all yarns are straight and penetrate each other
at crossings.  Then, for few steps, for any penetration detected between two fibers belonging to
crossing yarns, we take as normal direction of contact a vertical direction oriented according to
the local crossing order prescribed by the weave pattern. This process is applied until yarns are
completely separated at crossings. Then, in a second stage, classical contact directions, depending
only on the local geometry of fibers, are considered, while forces and displacements applied on the
sides of the patch are progressively relaxed. At the end of this process, an equilibrium
configuration of the woven structure after weaving is obtained.

\subsection{Application to the studied patches}

Two patches of fabric, made with the same yarns, but according to two different weave patterns --a
plain weave and a twill weave--, have been considered for the results presented here. The main
features characterizing these patches are given in Table \ref{tab:1}. Each patch is made of 408
fibers, and about 80.000 contact elements are considered in the simulation.

\begin{table}
%
%
\begin{tabular}{p{6cm}r}
\hline\noalign{\smallskip}
Nb. of fibers weft yarns & 44 \\
Nb. of fibers warp yarns & 24 \\
Nb. of weft yarns & 6 \\
Nb. of warp yarns & 6 \\
Total nb. of fibers & 408 \\
Nb. of nodes & 35.000 \\
Nb. of dofs & 300.000\\
Nb. of contact elements & $\approx$ 80.000\\
\noalign{\smallskip}\hline\noalign{\smallskip}
\end{tabular}
\caption{Characteristics of the studied patches}
\label{tab:1} 
\end{table}

\begin{figure}[!ht]
  \begin{centering}
    \raisebox{1.5cm}{(a)}\includegraphics[scale=0.2]{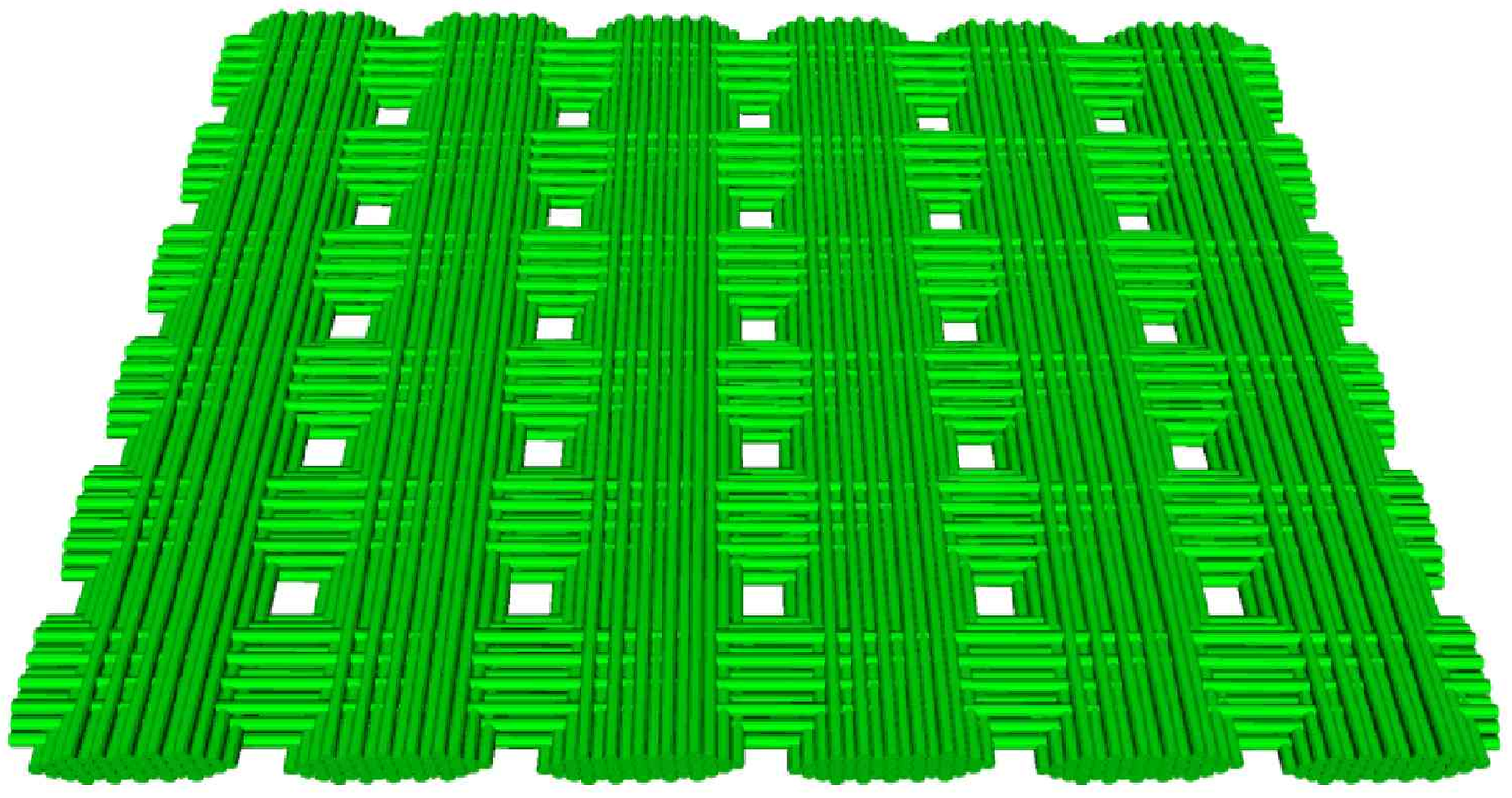} \\
    \raisebox{1.5cm}{(b)}\includegraphics[scale=0.2]{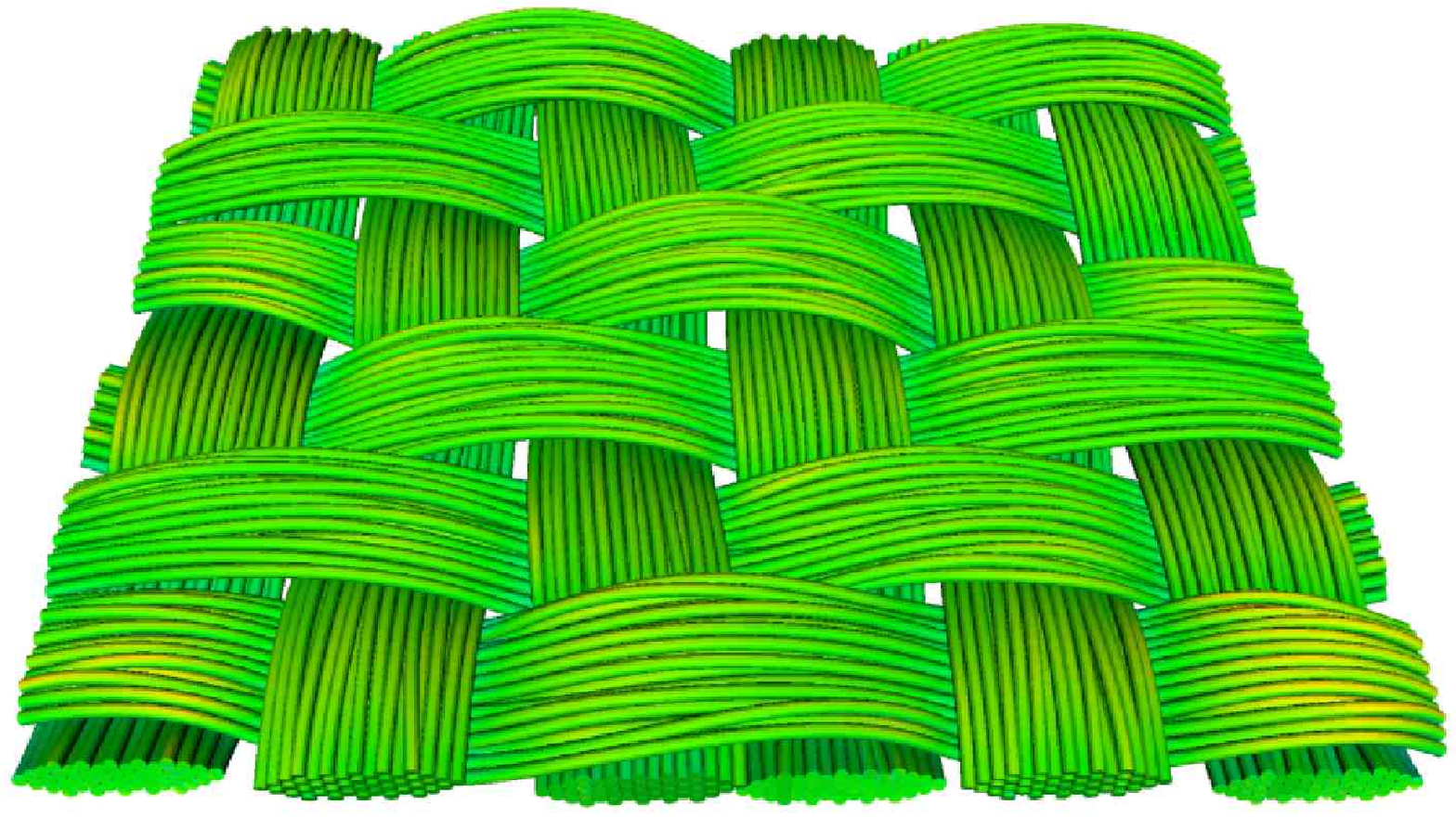}
    \raisebox{1.5cm}{(c)}\includegraphics[scale=0.2]{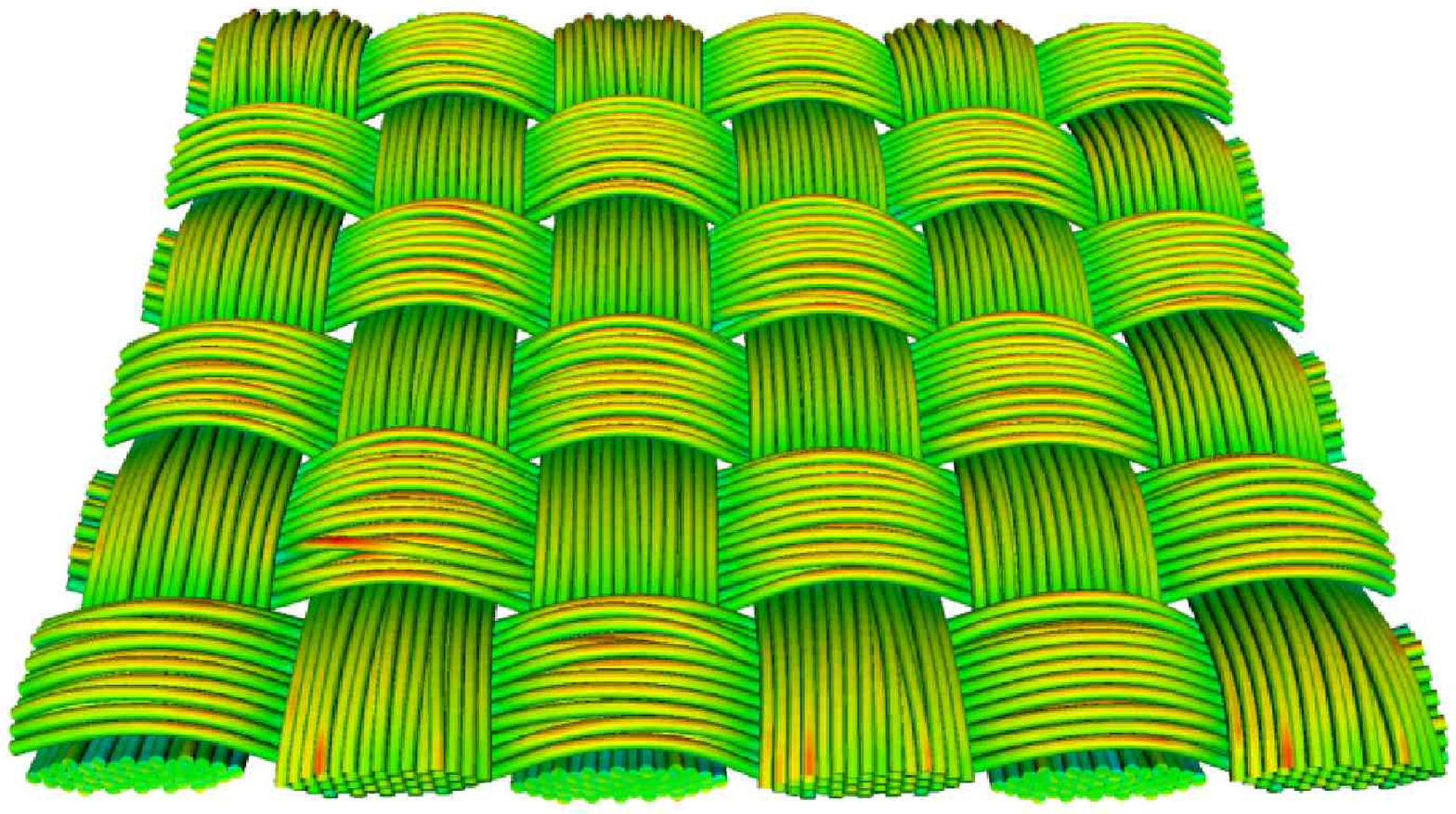}
    \caption{Simulation of the weaving process : configuration before weaving (a), computed
      configuration for plain weave (b) and twill weave (c)}
  \end{centering}
  \label{fig:Forming}     
\end{figure}

\begin{figure}[!ht]
  \begin{centering}
    \raisebox{0.5cm}{(a)}\includegraphics[scale=0.4]{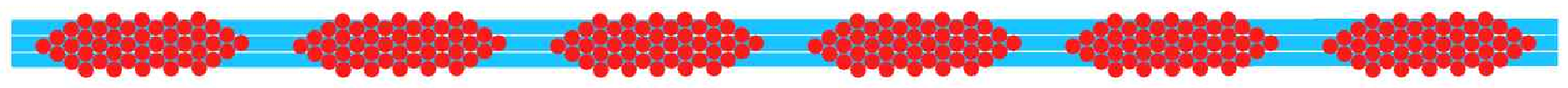}
    \raisebox{0.5cm}{(b)}\includegraphics[scale=0.4]{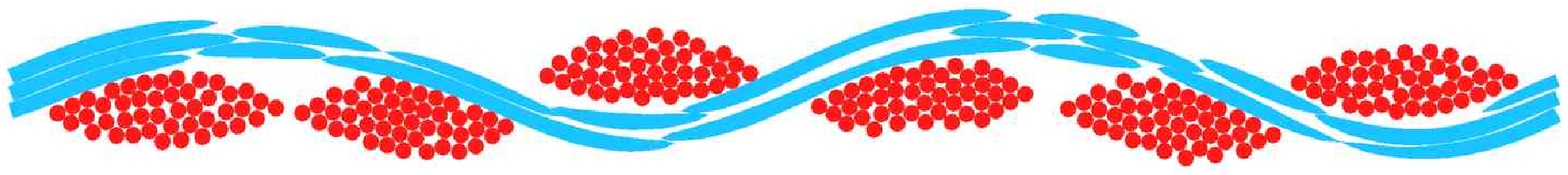}
    \raisebox{0.5cm}{(c)}\includegraphics[scale=0.4]{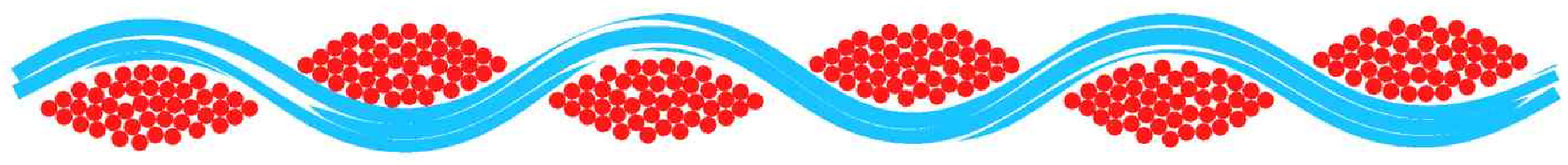}
  \end{centering}
  \caption{Cuts of configuration before weaving (a), and of computed configuration for a plain weave (b) and a twill weave
    (c) patches}
  \label{fig:CutForming}     
\end{figure}

\begin{figure}[!ht]
  \includegraphics[scale=0.4]{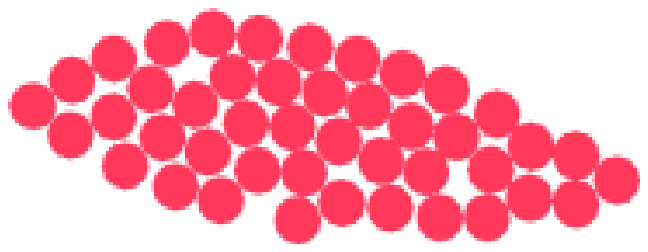}
  \includegraphics[scale=0.4]{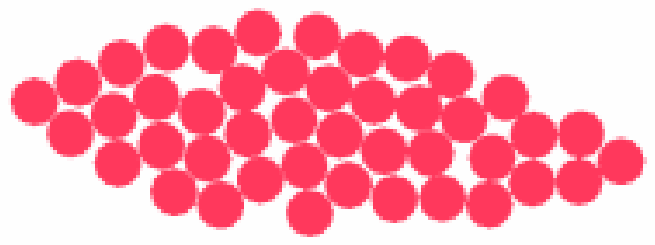}
  \includegraphics[scale=0.4]{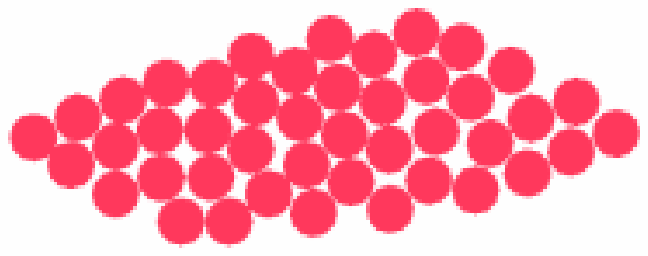}
  \includegraphics[scale=0.4]{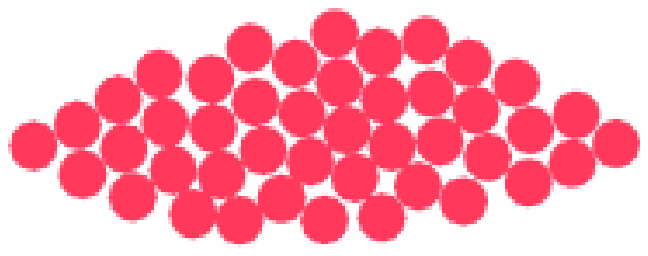}
  \caption{Variation of cross-section shapes along a warp yarn for twill weave}
  \label{fig:VarCrossSect}     
\end{figure}

Computed configurations for the two weavings are shown on Fig. \ref{fig:Forming}. Cuts of the two
computed initial configurations and details of these cuts (Fig. \ref{fig:CutForming}) show the
rearrangement of fibers and typical shapes of cross-sections obtained by the simulation. Whereas for
plain weaves shapes of yarn cross-sections are predominantly lenticular, for twill weave
cross-sections are more complex and vary along the yarn (Fig. \ref{fig:VarCrossSect}).

\subsection{Application of loading tests}

\begin{figure}[!ht]
\includegraphics[scale=0.22]{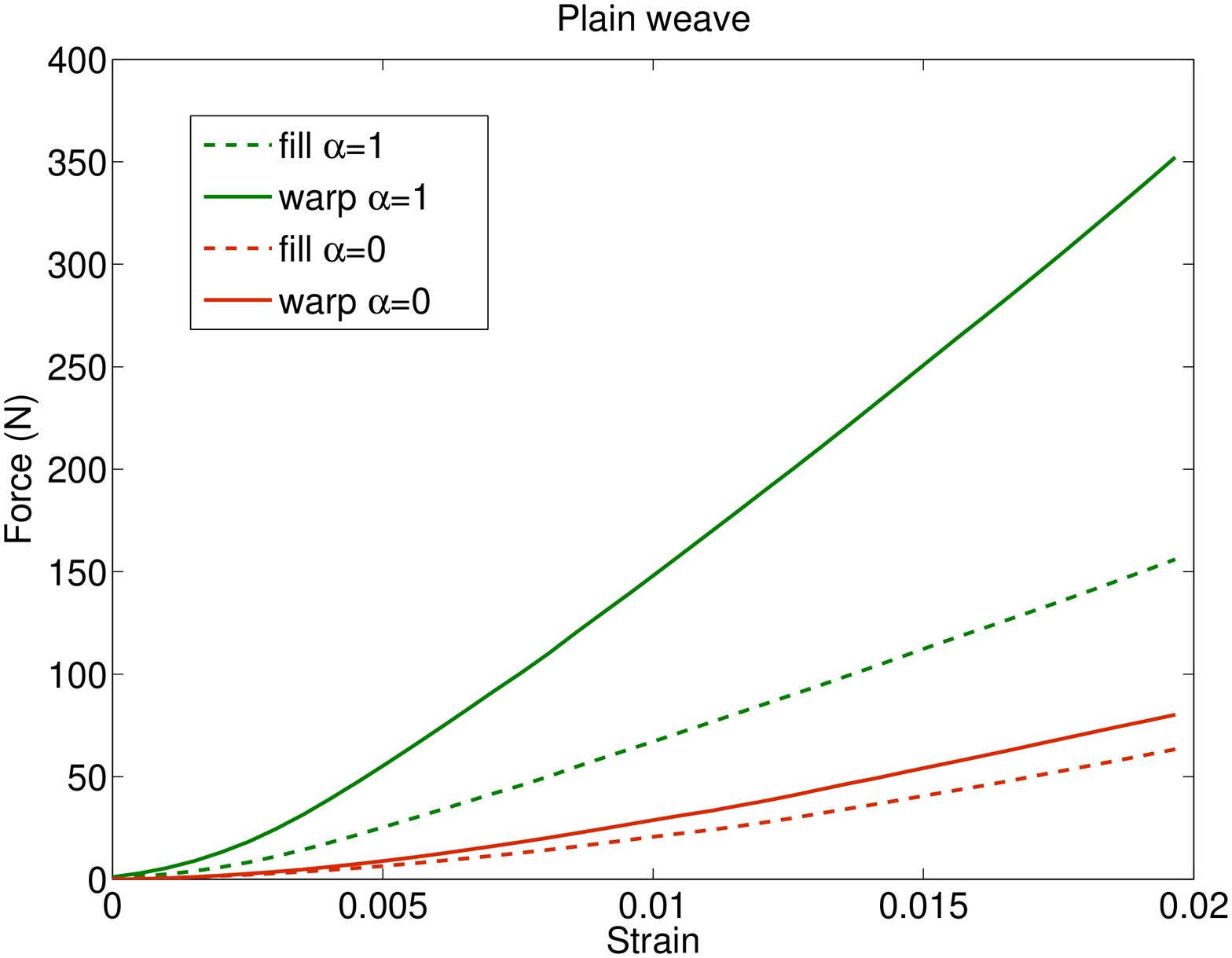}
\includegraphics[scale=0.22]{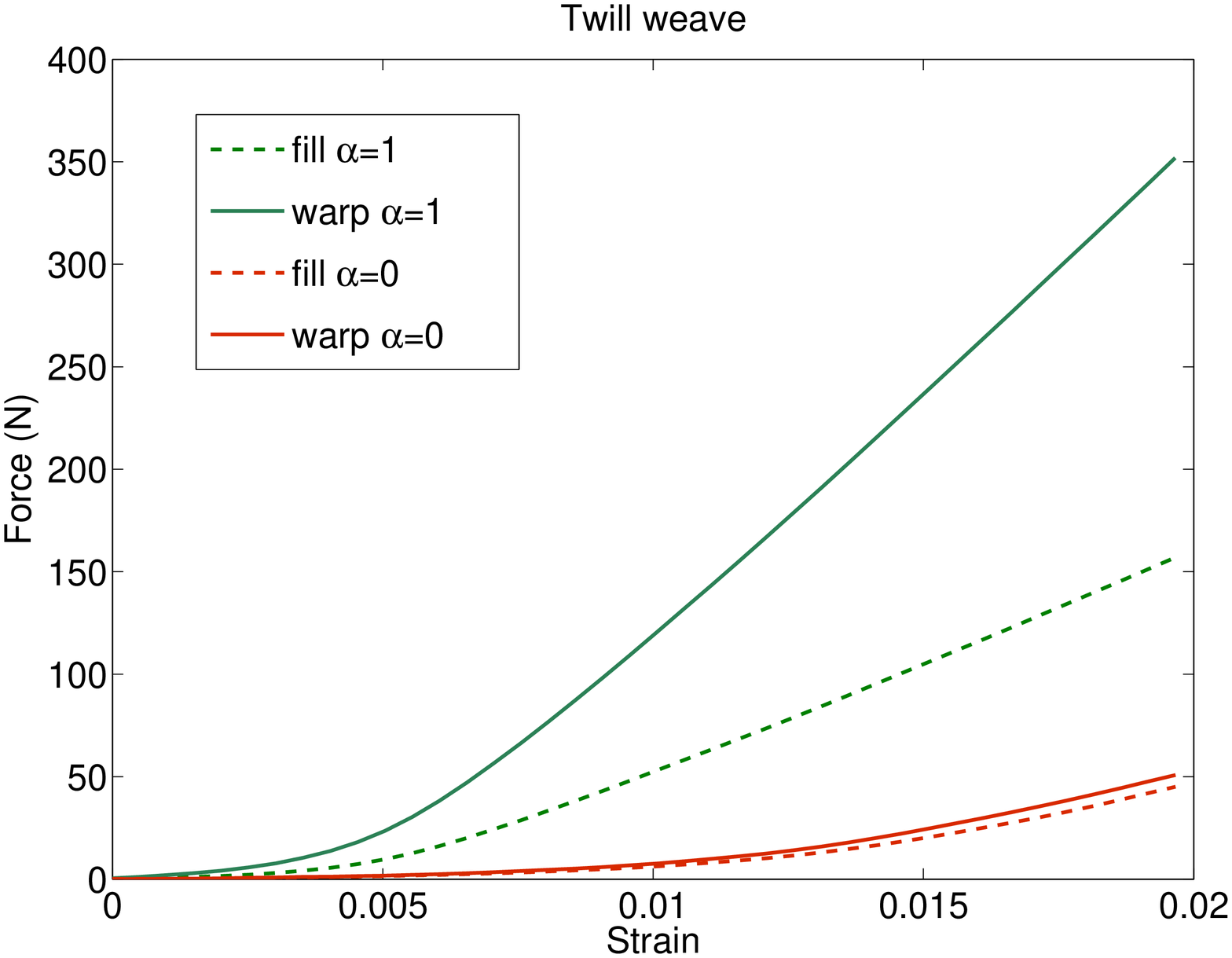}
\caption{Force/strain curves for biaxial tension tests for plain weave and twill weave}
\label{fig:Force}     
\end{figure}

Once the initial configuration has been computed, various loading cases can be simulated by applying
appropriate loadings (displacements or forces) to rigid bodies attached to each side of the patches.
Biaxial tension loadings are applied up to a $2 \%$ extension in the warp direction, with a ratio
$\alpha$, taken successively equal to 0 and 1, for the extension in the weft direction.  Typical
J-shape force/displacement curves are obtained (Fig. \ref{fig:Force}).  Nonlinear effects at the
start of the curves are related to the compaction of cross-sections as the tensile force increases.

\section{Strain measures at the scale of yarns}

The formulation of mesoscopic models to represent the behaviour of yarns requires to define measures
for both strains and stresses at the scale of yarn. The definition of a stress measure is difficult
because stresses within yarns are of two different kinds : continuous stresses inside yarns, and
discrete contact-friction interactions between fibers.

\begin{figure}[!ht]

\includegraphics[scale=0.3]{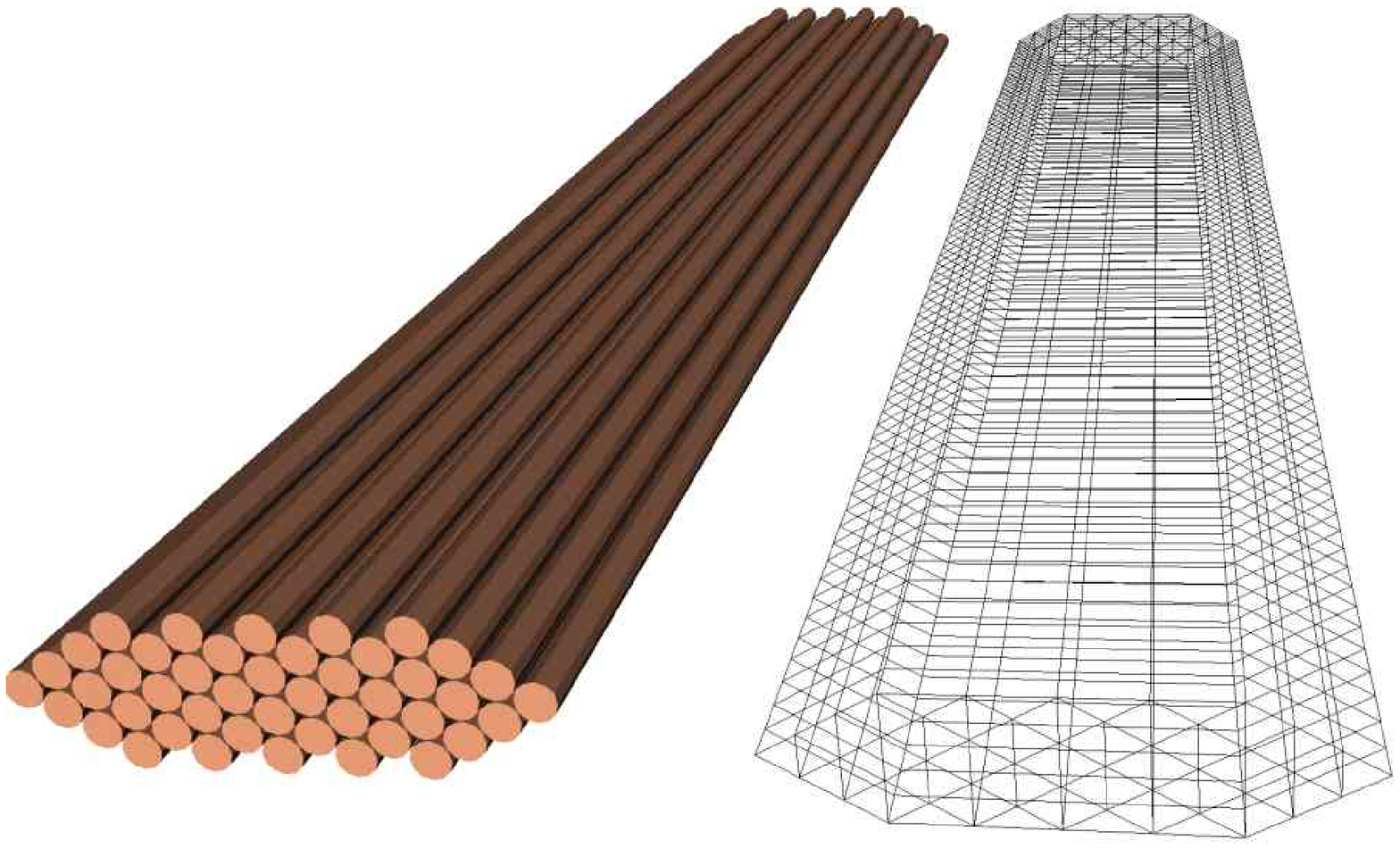}
\includegraphics[scale=0.3]{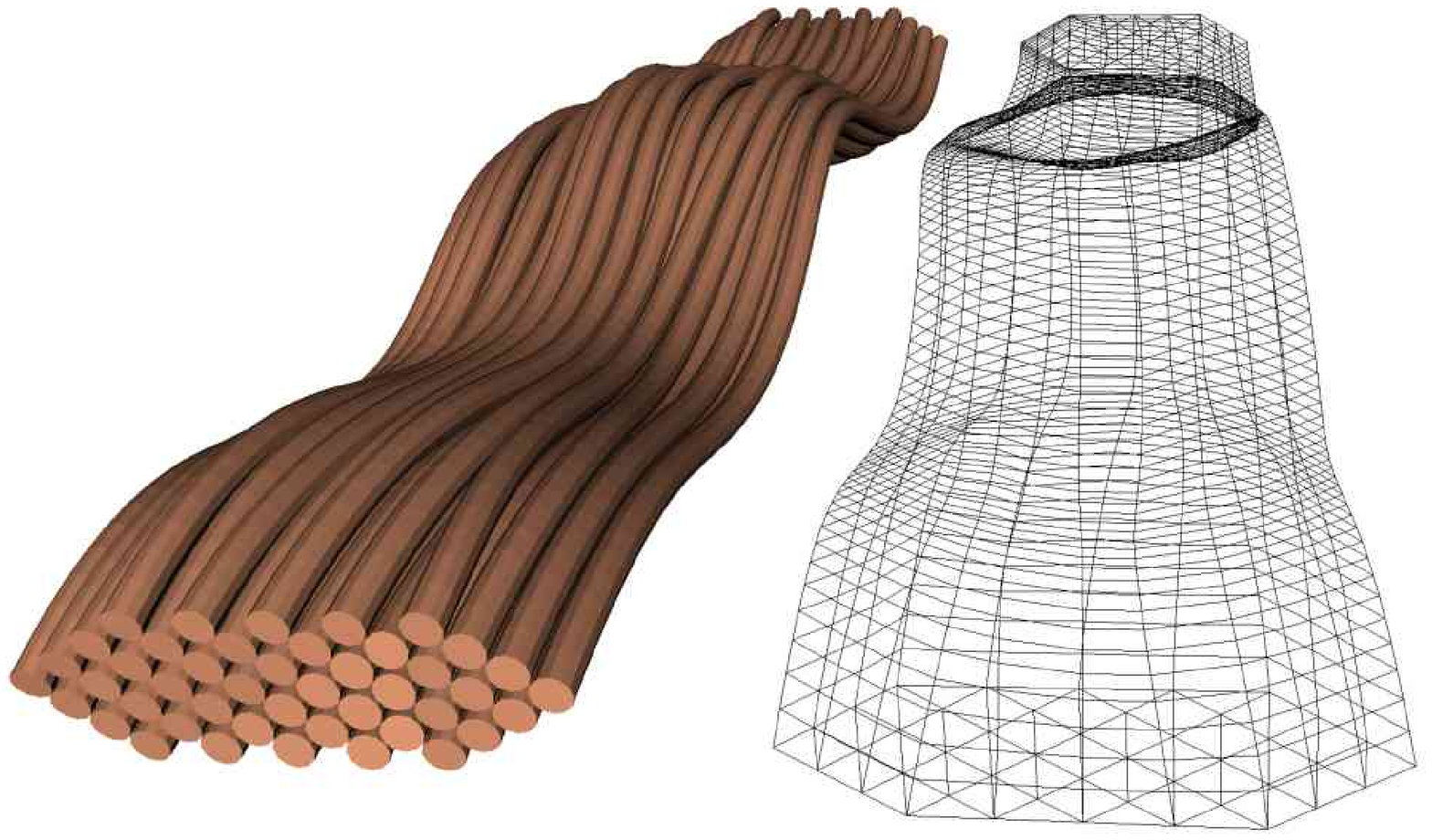}
%
\caption{Fibers of a yarn, and corresponding 3D meshes in two different configurations}
\label{fig:MeshYarn}     
\end{figure}

As a first step, the definition of a strain measure at the scale of yarn is easier. To approximate
such a quantity, we propose to consider each yarn as a continuum, and to compute, as
post-processing, Green-Lagrange strain tensors using a 3D finite element mesh based on the nodes
defined on fibers (Fig.\ref{fig:MeshYarn}). Once the mesh connectivity has been defined for the
initial configuration, nodal displacements and strain tensors can be easily derived. It is also
possible to consider strains generated by displacements only between two given loading increments.

\begin{figure}[!ht]
\includegraphics[scale=0.3]{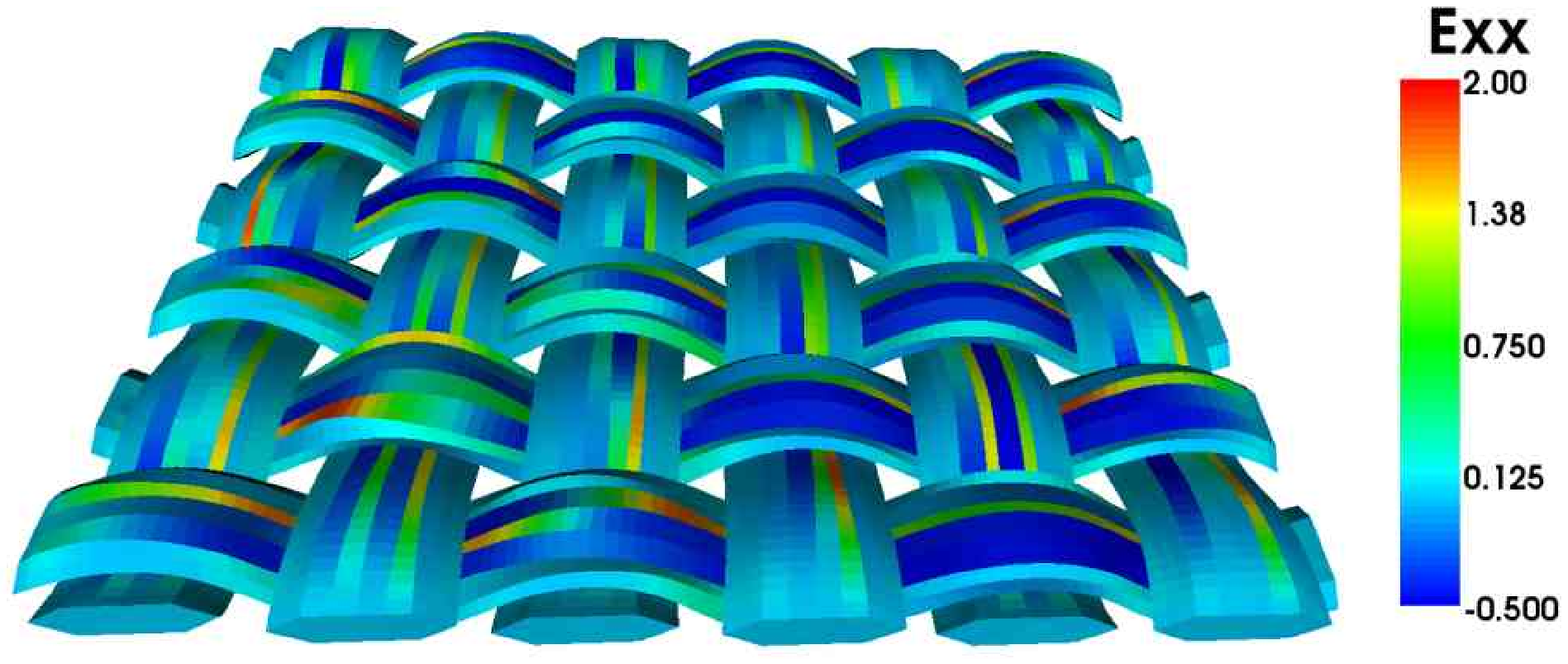}
\includegraphics[scale=0.3]{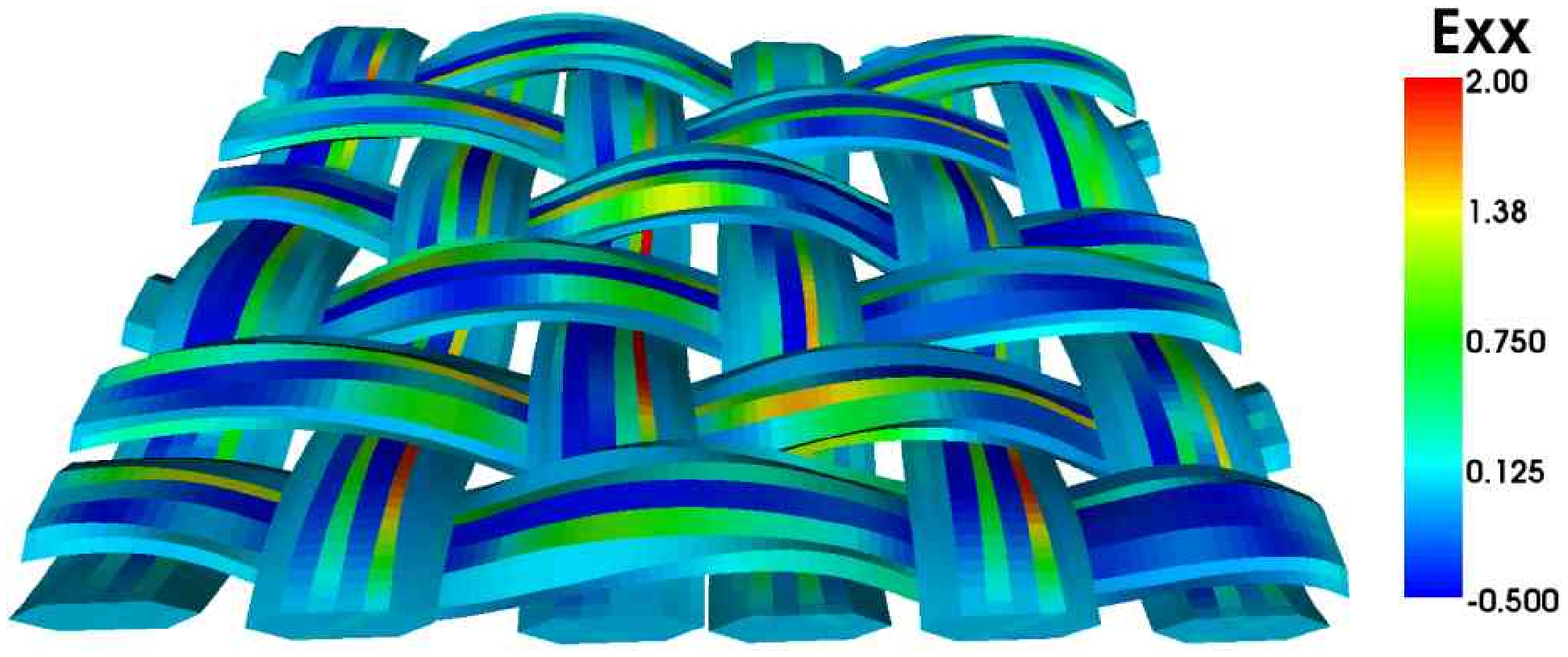}
\caption{Horizontal strains in yarns generated by the simulation of the weaving process}
\label{fig:Exx}     
\end{figure}

\begin{figure}[!ht]
\includegraphics[scale=0.15]{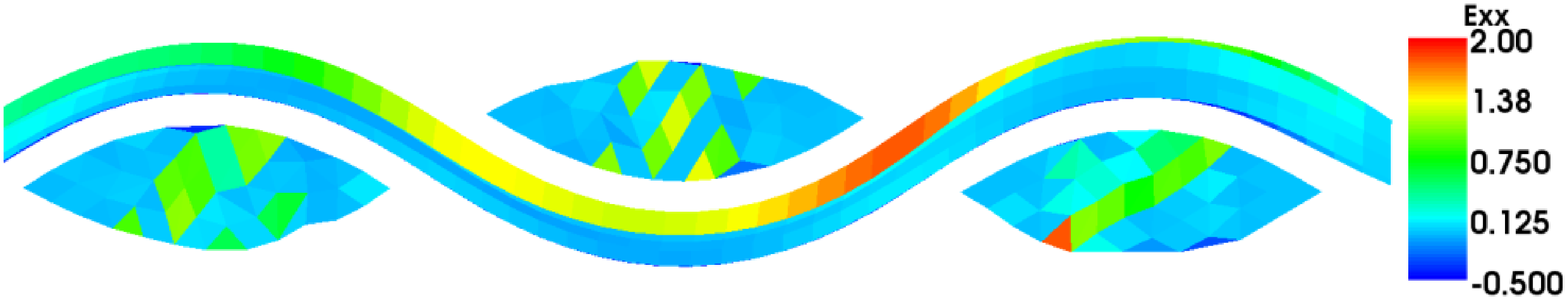}
\includegraphics[scale=0.15]{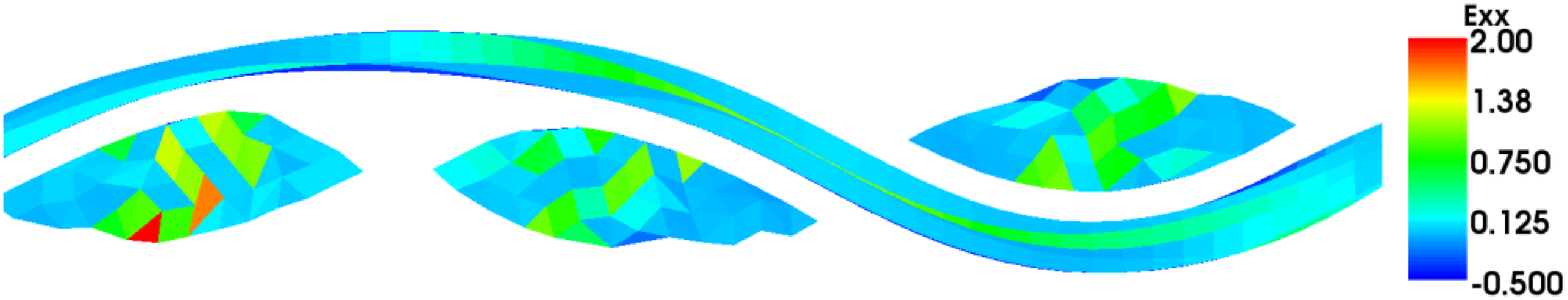}
\caption{Cuts of horizontal strains in yarns generated by the simulation of the weaving process}
\label{fig:ExxCut}     
\end{figure}

\begin{figure}[!ht]
\includegraphics[scale=0.15]{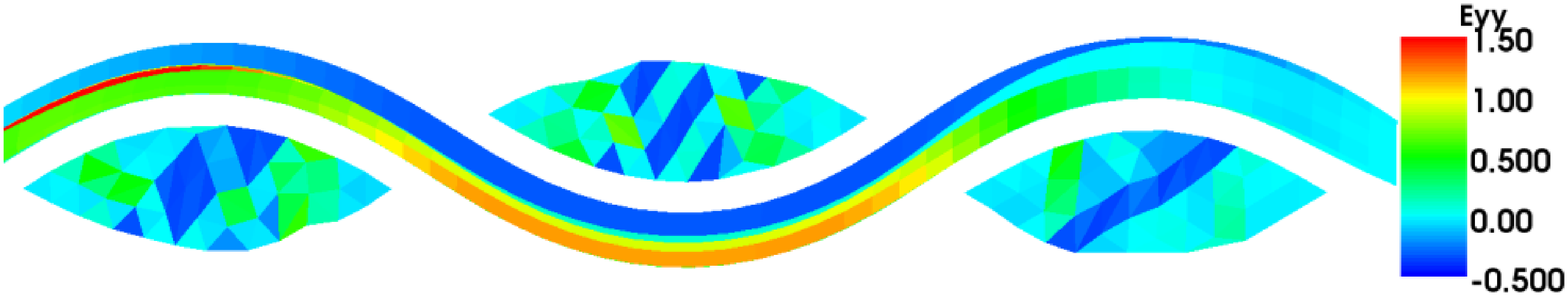}
\includegraphics[scale=0.15]{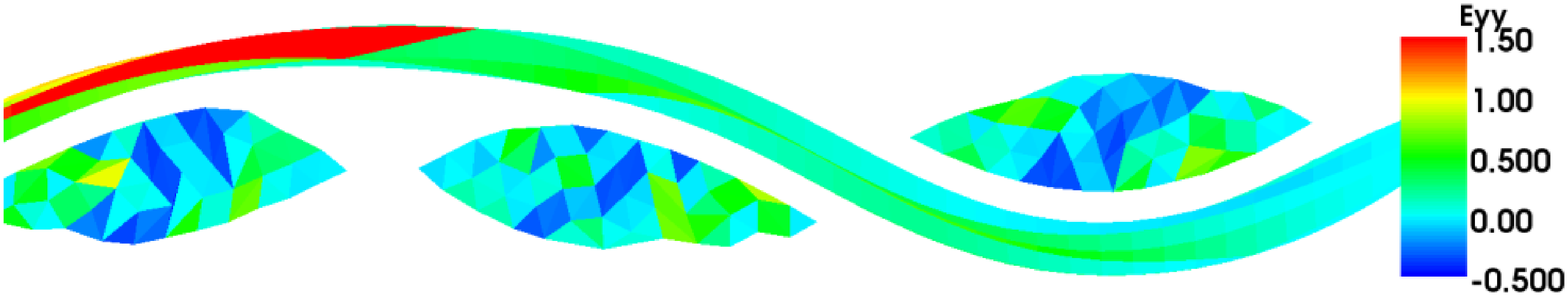}
\caption{Cuts of vertical strains in yarns generated by the simulation of the weaving process}
\label{fig:EyyCut}     
\end{figure}

The plotting of horizontal strains due to the initial forming (Fig. \ref{fig:Exx}) shows strong
inhomogeneities. Zones looking like diagonal shear bands can be observed on the cuts of strains in both
horizontal and vertical directions (Fig. \ref{fig:ExxCut} and \ref{fig:EyyCut}).

\begin{figure}[!ht]

\raisebox{1.cm}{(a) }\includegraphics[scale=0.48]{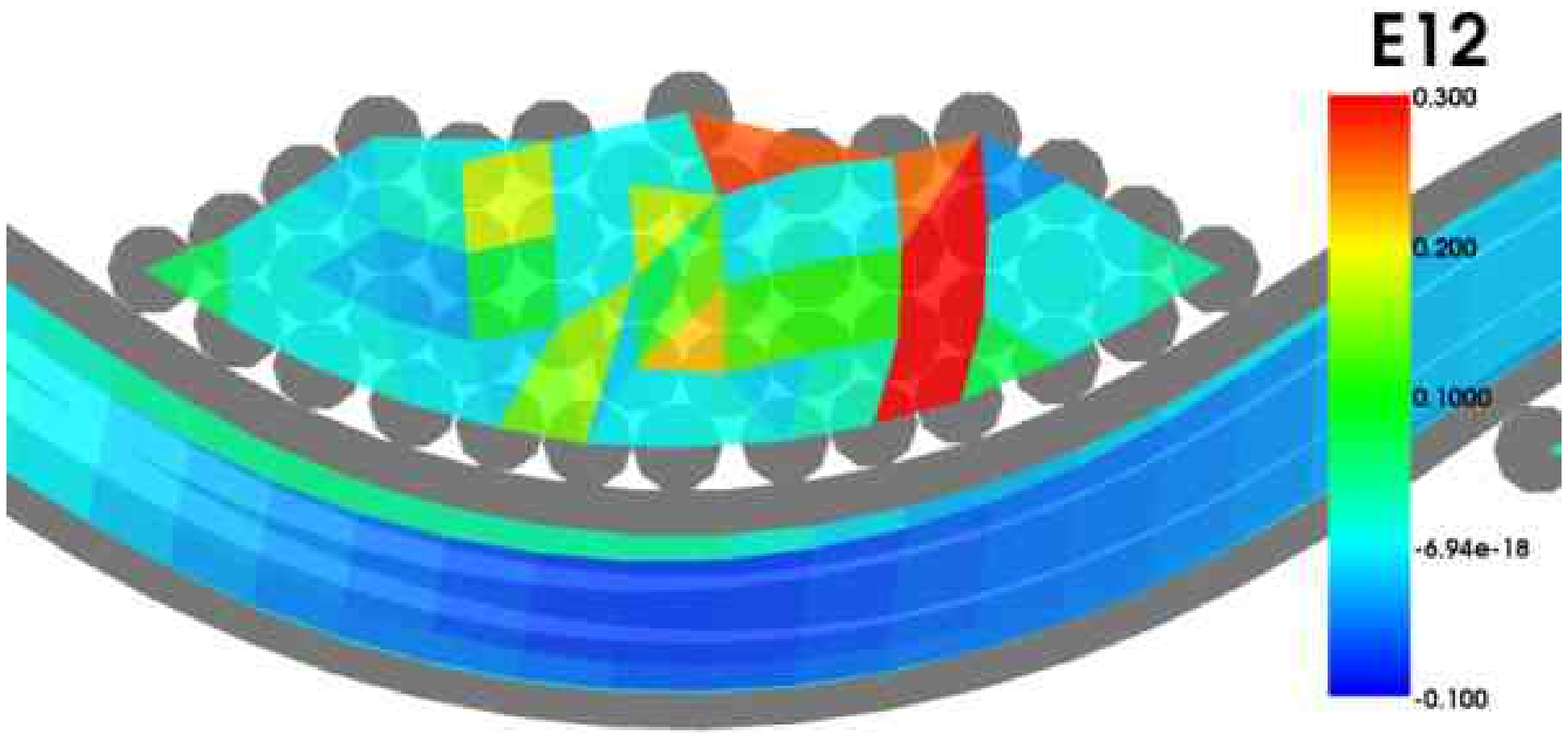}
\raisebox{1.cm}{(b) }\includegraphics[scale=0.48]{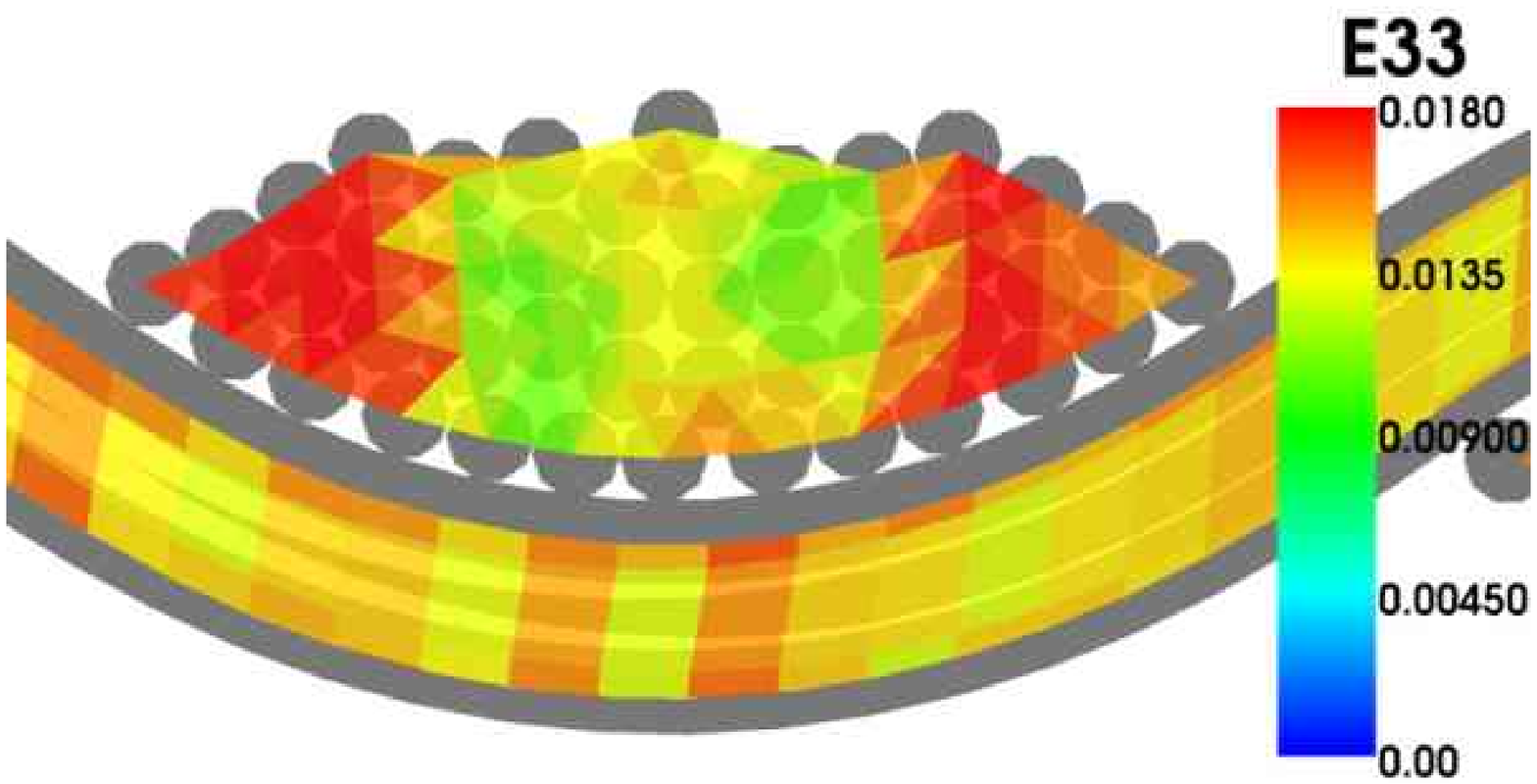}
%
\caption{Shear strains (a) and axial strains (b) in yarns generated by equibiaxial tension}
\label{fig:CutAxialStrains}     
\end{figure}

Similar inhomogeneities can be observed for strains induced by the equibiaxial tension (Fig.
\ref{fig:CutAxialStrains}). In this case, the axial strain is twice higher on the sides of the yarn
than in the center. The appearance of shear bands shows a rearrangement of fibers in the center of
the yarn. This rearrangement reduces the heigth of the cross-section, allowing fibers to undergo
lower extensions.

Inhomogeneities revealed by this post-processing are of first importance since they raise the
question of using continuous models to describe the behaviour of yarns.

\section{Conclusion}

Models and algorithms developped for the simulation of the mechanical behaviour of entangled media
can be applied for the modelling of woven structures. Patches of fabric made of few hundreds of
fibers can be considered by the model. The simulation of the weaving process provide accurate
geometrical descriptions of both yarns trajectories and varying shapes of yarns cross-sections. The
mechanical response of the fabric can then be characterized by simulating typical loading tests. The
computation of 3D strains at the scale of yarns reveals strong inhomogeneities and raises the
question of the validity of considering yarns as continuums.

\bibliographystyle{spmpsci}
\bibliography{durville}
\end{document}